\documentclass[
 reprint,
 amsmath,amssymb,
 aps,
 superscriptaddress,
prb,
]{revtex4-2}

\usepackage[colorlinks,allcolors=blue,urlcolor=blue]{hyperref}
\usepackage{graphicx}
\usepackage{dcolumn}
\usepackage{bm}
\usepackage{xcolor}
\usepackage{physics}
\usepackage{gensymb}

\newcommand{\Rp}{\mathbf{R_p}}
\newcommand{\Rel}{\mathbf{R_e}}
\newcommand{\Velph}{\hat{V}_{\mathrm{e}-\mathrm{ph}}}

\begin{document}

\preprint{APS/123-QED}

\title{Predicting Electron-Phonon Coupling and Electronic Transport \\ at the Moir\'e Scale in Twisted Bilayer Graphene}

\author{David J. Abramovitch}%
\affiliation{Department of Applied Physics and Materials Science, and Department of Physics, California Institute of Technology, Pasadena, California 91125}
\author{Marco Bernardi}
\email{bmarco@caltech.edu}
\affiliation{Department of Applied Physics and Materials Science, and Department of Physics, California Institute of Technology, Pasadena, California 91125}

\begin{abstract} 
First-principles calculations can accurately describe electron-phonon (e-ph) interactions and electronic transport in a wide range of materials, but are currently limited 
to unit cells with up to $\sim$100 atoms due to computational cost.
Here, we develop an atomistic electronic potential with Holstein- and Peierls-like terms for modeling e–ph interactions and phonon-limited electronic transport that enables the study of moiré systems with thousands of atoms per unit cell.
This method can accurately reproduce first-principles e–ph coupling and resistivity in graphene and large-angle twisted bilayer graphene (TBG). 
Using this approach, we study TBG over a range of twist angles down to 1.6\degree~ (5044-atom unit cell), and report the evolution of e-ph interactions and phonon-limited resistivity with twist angle. 
The predicted resistivity increases by two orders of magnitude between 13.2~\degree and 1.6\degree, driven by the progressive reduction of the electronic energy scale. 
Our calculations can predict key experimental trends in 2.0\degree and 1.6\degree~TBG, including the resistivity and its dependence on temperature and band filling. 
Our work establishes a scalable approach for quantitative studies of e–ph interactions and transport in moir\'e materials and other systems with previously inaccessible length scales.
\end{abstract}

\maketitle

\begin{figure*}
    \centering
    \includegraphics[width=\textwidth]{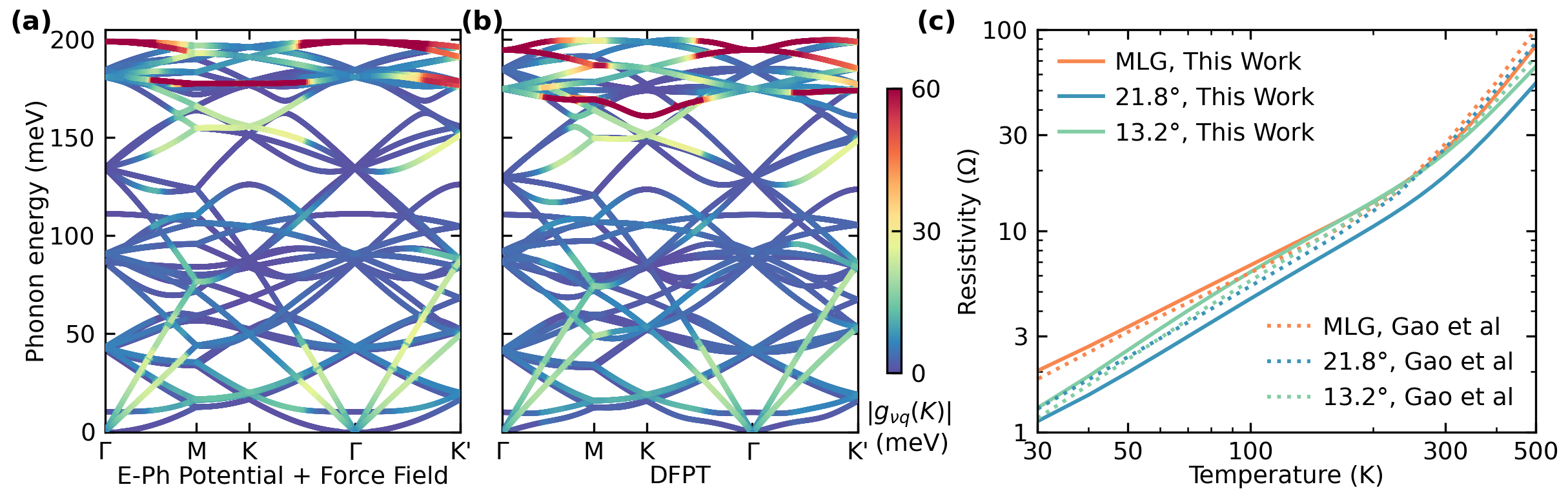}
    \caption{(a) E-ph coupling computed in this work, $g_{\nu}(\mathbf{k} \!=\! K, \mathbf{q}) = (\frac{1}{N_b}\sum_{nm} |g_{nm\nu}(\mathbf{k} = K, \mathbf{q})|^2)^{1/2}$ averaged over Dirac-cone bands, plotted on the phonon dispersion for 21.8\degree~TBG. (b) Same plot as in (a), but with phonons and e-ph coupling from DFPT. (c) Resistivity in MLG, calculated with our approach and from DFPT e-ph coupling~\cite{gao_firstprinciples_2024}, in both cases using the RTA. }
    \label{fig:epc_model_vs_dfpt}
\end{figure*}

\begin{figure}[b]
    \centering
    \includegraphics[width=0.9\linewidth]{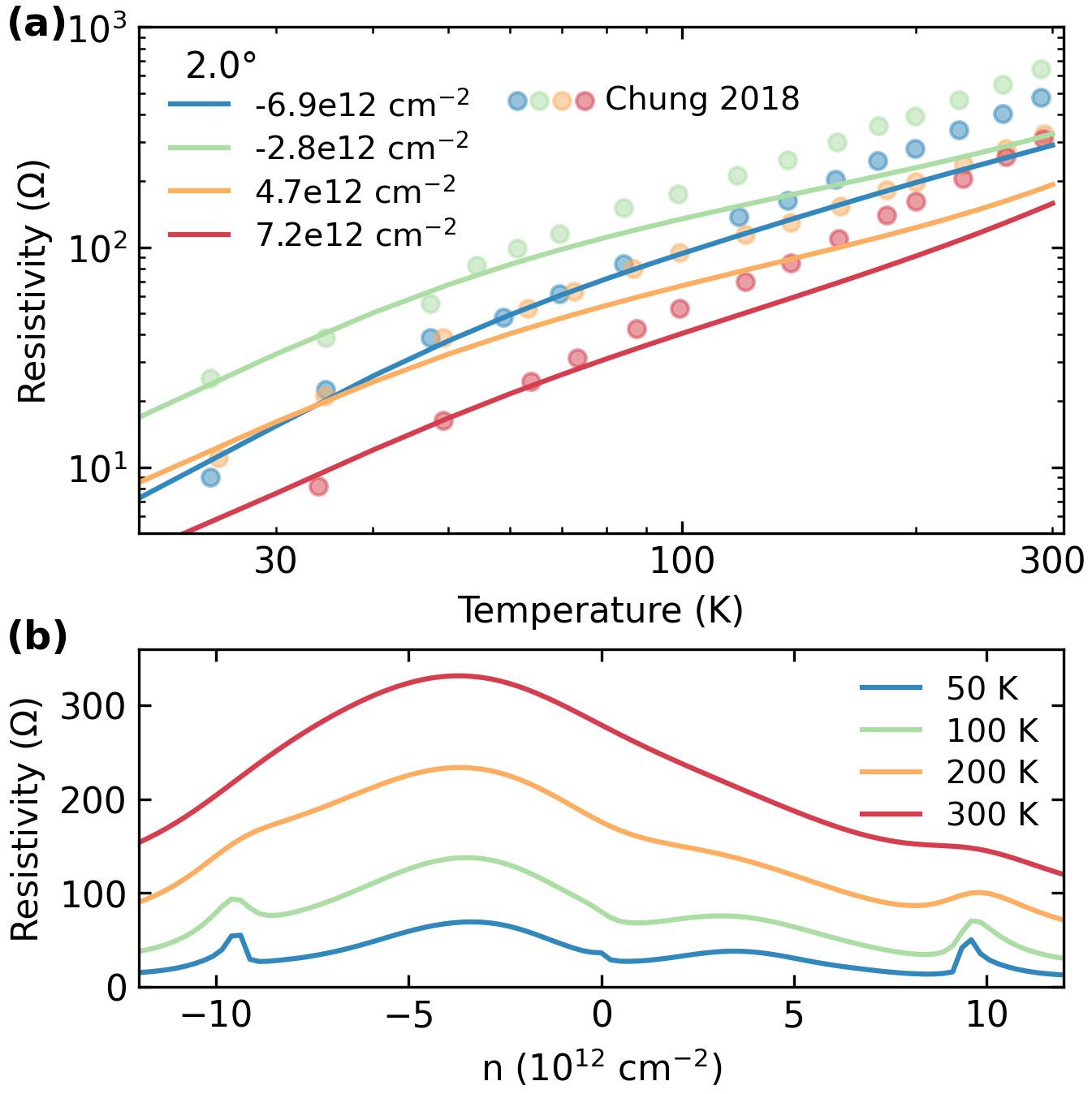}
    \caption{(a) Temperature dependent resistivity of 2.0\degree~TBG computed for several electronic fillings (solid lines) and compared to experiments (symbols)~\cite{chung_transport_2018}. (b) Filling-dependent resistivity in 2.0\degree~TBG at several temperatures. The moir\'e bands are completely filled / empty at $n \!=\! \pm 9.3 \times 10^{12}$ cm$^{-2}$.}
    \label{fig:2deg}
\end{figure}

\begin{figure}[t]
    \centering
    \includegraphics[width=0.9\linewidth]{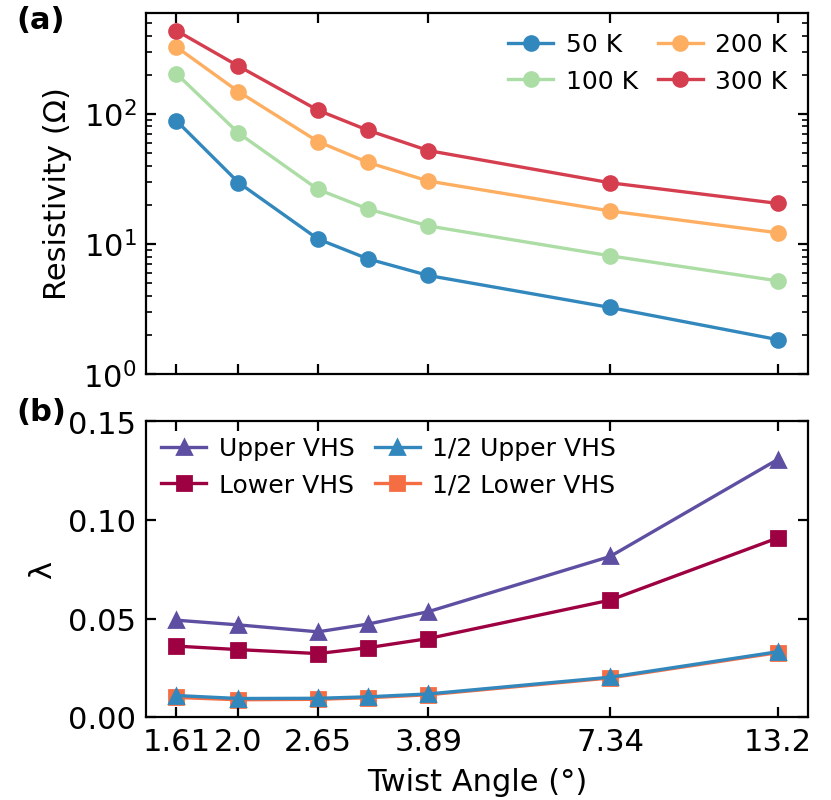}
    \caption{(a) Twist angle dependence of the resistivity at several temperatures for a Fermi energy located halfway to the upper VHS. (b) Twist angle dependence of the e-ph coupling strength $\lambda$ at four fillings. }
    \label{fig:twist-angle-all}
\end{figure}

\begin{figure*}
    \centering
    \includegraphics[width=0.91\textwidth]{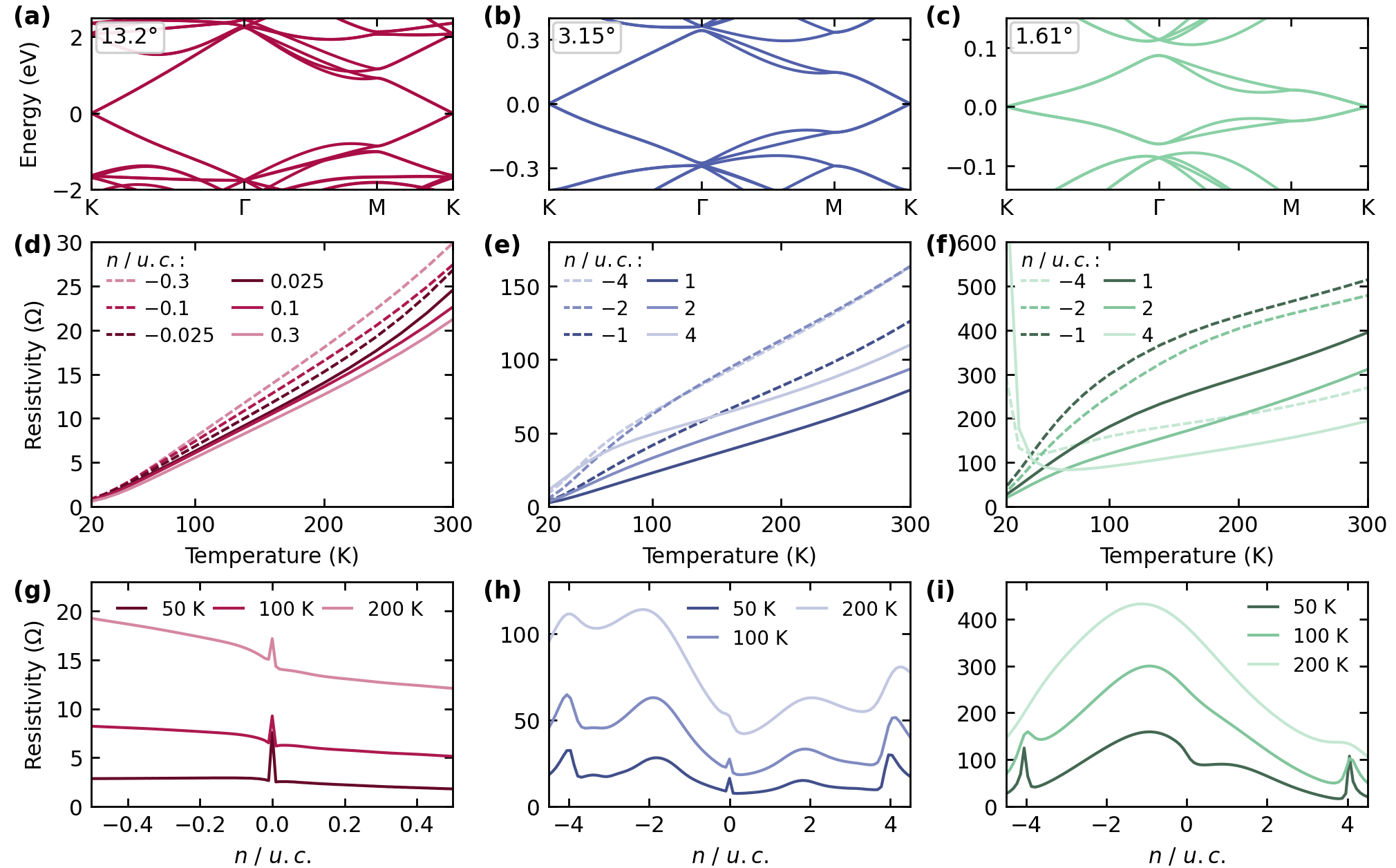}
    \caption{Calculated band structure of (a) 13.2\degree, (b) 3.15\degree, and (c) 1.61\degree TBG, respectively. For the same sequence of twist angles, we show the temperature dependent resistivity at several electronic fillings in panels (d)-(f), and the resistivity as a function of electron filling at several temperatures in panels (g)-(i), respectively.}
    \label{fig:transport-twist-angle}
\end{figure*}

\vspace{-10pt}
First-principles calculations of electronic transport achieve predictive accuracy in classes of materials ranging from metals~\cite{Mustafa} and semiconductors~\cite{wuli2015, zhou_abinitio_2016, Liu-first, ponce2018, naph, yliu, ponce2020first, chang_bandlike_2025} to complex oxides~\cite{STO-tdep, luo_first_2025, lihm_beyond_2026} and correlated materials~\cite{deng_transport_2016, David_prm, abramovitch_respective_2024}. 
These calculations combine first-principles electronic structure, phonons, and electron-phonon (e-ph) interactions~\cite{mahan2011condensed, bernardi_first_2016, giustino_electron_phonon_2017}, building on the framework of density functional theory (DFT) and linear-response DFT~\cite{baroni_phonons_2001}. 
Studies of electronic transport in graphene and related systems illustrate these developments: following early analytic models~\cite{hwang_acoustic_2008, dassarma_electronic_2011, kaasbjerg_unraveling_2012}, first-principles calculations have studied band renormalization and electron scattering rates~\cite{park_velocity_2007, park_firstprinciples_2009},
temperature dependent resistivity~\cite{park_electronphonon_2014, sohier_phononlimited_2014}, and more recently Hall mobility~\cite{desai_magnetotransport_2021}, twisted bilayer structures~\cite{gao_firstprinciples_2024}, and coupled electron and phonon dynamics~\cite{tong_toward_2021, yao_advancing_2025}. 
\\
\indent
Electronic transport in twisted bilayer graphene (TBG) has been studied intensely as the emergent moir\'e physics leads to unconventional behaviors. Experimental studies have reported a large resistivity in small-angle TBG over a range of temperatures, electron fillings, and twist angles~\cite{chung_transport_2018, polshyn_large_2019,cao_strange_2020, jaoui_quantumcritical_2022}. Measurements of $\sim$2$\degree$ TBG have shown power-law resistivity and attributed this behavior to acoustic phonon scattering~\cite{chung_transport_2018}. Near the magic angle, observations of  $T$-linear resistivity~\cite{polshyn_large_2019,cao_strange_2020, jaoui_quantumcritical_2022} have been explained in terms of mechanisms including Planckian scattering and strange metallicity~\cite{cao_strange_2020, jaoui_quantumcritical_2022}, e-ph scattering reminiscent of monolayer graphene (MLG)~\cite{polshyn_large_2019, wu_phonon_2019, sharma_carrier_2021}, and dominant electron-electron scattering~\cite{shilov_interaction_2025}. 
\\
\indent 
The electronic structure of TBG is challenging to describe~\cite{carr_electronic_2020} due to the combination of large system size, sensitivity to the atomic structure~\cite{pathak_accurate_2022, krongchon_registry_2023}, and correlation effects~\cite{cao_unconventional_2018, cao_correlated_2018, choi_electronic_2019, choi_interaction_2021, goodwin_twist_2019, xie_nature_2020}. Electronic transport poses additional challenges, including the sensitivity of e-ph interactions to the electronic structure and phonons, and the need to account for a large number of scattering processes. 
Although theoretical studies in the Boltzmann picture have proposed phonon-limited transport as a viable explanation for the observed resistivity, these works rely on simplifying approximations such as Dirac electronic states~\cite{wu_phonon_2019,das_sarma_strange_2022}, heuristic e-ph coupling constants~\cite{wu_phonon_2019,sharma_carrier_2021,das_sarma_strange_2022}, and the inclusion of only intralayer~\cite{sharma_carrier_2021,wu_phonon_2019,das_sarma_strange_2022} or breathing and flexural phonons~\cite{ray_electronphonon_2016}.  
\\
\indent 
However, a treatment of e–ph interactions and electronic transport at the moiré scale with first-principles accuracy is still lacking.  
Despite the development of scalable approaches for e-ph interactions~\cite{samsonidze_accelerated_2018, ganose_efficient_2021, luo_datadriven_2024, zhong_accelerating_2024, li_deep_2024}, first-principles e-ph calculations are currently limited to systems with 50$-$100 atoms in the unit cell~\cite{ gao_firstprinciples_2024, chang_bandlike_2025}. 
A relevant example is a recent study of e-ph interactions and transport in large-angle TBG, which showed that layer-breathing phonons can substantially modify the temperature dependence of the resistivity from the $T$-linear behavior characteristic of MLG~\cite{gao_firstprinciples_2024}. Extending these calculations to low twist angles by brute force is not possible due to the computational cost of linear-response DFT. 
\\ 
\indent 
In this Letter, we show scalable calculations of \mbox{e-ph} interactions and phonon-limited resistivity that can reach the moir\'e scale while retaining first-principles accuracy. 
This is achieved through a generalized Holstein-Peirls ansatz for the e-ph interactions, parametrized using DFT and linear-response DFT. 
This method relies on a small number of short-range terms that depend on the local atomic positions, and can accurately reproduce first-principles e-ph interactions and resistivity in MLG and large-angle TBG, used here as benchmarks. 
Our \mbox{approach} enables the study of e-ph coupling and electronic transport in TBG for twist angles as small as 1.61\degree, corresponding to moir\'e unit cells with thousands of atoms. 
Our predicted resistivity in TBG, computed in the Boltzmann picture, increases substantially for decreasing twist angles, mainly driven by a corresponding decrease in the electron band velocity. 
Our calculations achieve very good agreement with the experimental resistivity at low twist angles, indicating that the resistivity of TBG is largely phonon-limited down to $\sim$2\degree~twist angles despite the substantial interlayer electronic coupling at such low angles.
These findings sheds light on the microscopic mechanisms of electronic transport in TBG at a previously inaccessible length scale, and our approach provides a promising route to study transport phenomena in emerging moir\'e material platforms.\\
\indent  
\textit{Electron-Phonon Coupling in TBG.\textemdash}
The e-ph coupling $g_{mn\nu}(\mathbf{k},\mathbf{q})$ is often calculated as a Fourier transform of local real-space contributions~\cite{giustino_electron_phonon_2007, agapito_abinitio_2018, bernardi_first_2016, giustino_electron_phonon_2017}:
\vspace{-7pt}
\begin{multline}
    g_{mn\nu}(\mathbf{k},\mathbf{q}) =  
    \frac{1}{N_e}\sum_{ija}\sum_{\mathbf{R}_e\mathbf{R}_p} e^{i(\mathbf{k}\cdot\mathbf{R}_e + \mathbf{q}\cdot\mathbf{R}_p)} \sqrt{\frac{\hbar}{2M_a\omega_{\nu\mathbf{q}}}} 
    \\\times \mathbf{e}^{a\mu}_{\nu\mathbf{q}} U_{mi,\mathbf{k}+\mathbf{q}}\bra{i\mathbf{0}}\partial_{u_{a\mu\Rp}} V \ket{j\mathbf{R}_e} U^\dagger_{nj,\mathbf{k}}
    \label{eq:el-ph-fourier}
\end{multline}
where $m$ and $n$ index electronic bands, $i$ and $j$ index local orbitals, $a$ indexes an atom with mass $M_a$, $\mu$ is a Cartesian direction, $\mathbf{k}$ and $\mathbf{q}$ are crystal momenta of electrons and phonons respectively, $\mathbf{R}_e$ and $\mathbf{R}_p$ are corresponding lattice vectors, $U_{nj,\mathbf{k}}$ transforms from the orbital to the Bloch basis, $\omega_{\nu\mathbf{q}}$ is the phonon frequency, and $\mathbf{e}^{a \mu}_{\nu \mathbf{q}}$ is the phonon eigenvector. The change in Kohn-Sham potential with respect to atomic motions, $\partial_{u_{a\mu\Rp}} V$, is typically computed using density functional perturbation theory (DFPT). When the e-ph interactions are short-ranged, the sum converges with only a few terms~\cite{giustino_electron_phonon_2007, agapito_abinitio_2018}.
\\
\indent
To obtain a scalable approach that can reach the moir\'e scale, we formulate a tight-binding–style approximation where the e-ph coupling is generated by an electronic potential, $\Velph$, written as an operator in the atomic-orbital basis, with onsite and hopping matrix elements depending on a small number of atomic positions:
\vspace{-7pt}
\begin{multline}
      \Velph(\{\mathbf{r}_{a\mathbf{R}}\}) =  \sum_{i}\epsilon_{i}(\{\mathbf{r}_{a\mathbf{R}}\})c_{i\mathbf{0}}^\dagger c_{i\mathbf{0}} + \\ \sum_{\substack{i, j,\Rel}}(t_{i j\Rel}(\{\mathbf{r}_{a\mathbf{R}}\})c_{i\mathbf{0}}^\dagger c_{j\Rel} + h.c.).
\end{multline}
Here, $c_{i\Rel}^\dagger$ creates an electron in a localized orbital $\phi_{i\Rel}(\mathbf{r}) \!=\! \varphi_i(\mathbf{r} - \mathbf{\tau}_i - \Rel)$, and the onsite energy $\epsilon_{i}(\{\mathbf{r}_{a\mathbf{R}}\})$ and hopping matrix elements $t_{ij\Rel}(\{\mathbf{r}_{a\mathbf{R}}\})$ are differentiable functions of the atom positions $\{\mathbf{r}_{a\mathbf{R}}\}$. The real-space e-ph couplings are then calculated as
\begin{equation}
    \bra{i\mathbf{0}}\partial_{u_{a\mu\Rp}} \Velph \ket{j\Rel} = \partial_{u_{a\mu\Rp}} \left[\epsilon_{i}\delta_{i\mathbf{0},j\Rel} +  t_{ij\Rel }\right].
    \label{eq:coupling-from-V}
\end{equation}
\indent
In our approach (see end matter), the onsite terms in $\Velph$ are Holstein-like~\cite{holstein_studies_1959}, while the hopping terms are Peierls-like~\cite{barisic_tightbinding_1970} and include 1st-4th nearest intralayer neighbors and short-ranged interlayer hoppings. Both onsite and hopping terms depend on the positions of nearby atoms in both graphene layers. 
This provides a versatile ansatz for short-range e-ph coupling, which we parameterize from DFPT calculations; in polar materials, a long-range Fr\"olich term could also be included~\cite{frolich_electrons_1954}.
In combination with a tight-binding Hamiltonian~\cite{pathak_accurate_2022} for electronic structure and a force-field for lattice dynamics~\cite{perebeinos_valence_2009,krongchon_registry_2023}, this approach enables e-ph coupling and transport calculations in moir\'e systems with more than 5000 atoms. Our Holstein-Peirls ansatz differs from previous calculations of e-ph coupling in TBG based on derivatives of a Slater-Koster tight-binding Hamiltonian employed to study superconductivity and phonon linewidths~\cite{choi_strong_2018, choi_dichotomy_2021, mandal_phonon_2024}.\\ 
\indent
\textit{Large Angle TBG.\textemdash}
First, we benchmark our approach against DFPT-based e-ph calculations in large angle (21.8\degree)~TBG~\cite{gao_firstprinciples_2024}. Figures~\ref{fig:epc_model_vs_dfpt}(a) and \ref{fig:epc_model_vs_dfpt}(b) compare the e-ph coupling strength, color-coded on the phonon dispersion, obtained with our approach and with DFPT. We find very good agreement for both optical and acoustic modes. Importantly, the e-ph coupling is accurately reproduced for the longitudinal acoustic (LA), transverse acoustic (TA), and layer breathing (LB) modes, which play a key role in transport in MLG~\cite{park_electronphonon_2014} and TBG~\cite{gao_firstprinciples_2024}. The model also captures the momentum dependence of the coupling, including the reduced e-ph coupling for the LB mode at 20 meV near $\Gamma$ and the differences between inter- and intravalley couplings, $g(\mathrm{K},\mathrm{K})$ and $g(\mathrm{K},\mathrm{K}')$. 
\\
We additionally compute the temperature dependent resistivity in MLG and in two large-angle TBG structures (21.8\degree~and 13.2\degree). Figure~\ref{fig:epc_model_vs_dfpt}(c) compares results obtained from our approach with calculations using first-principles band structure, phonons, and e-ph coupling from Ref.~\cite{gao_firstprinciples_2024}, computed in both cases using the Boltzmann equation in the relaxation time approximation (RTA)~\cite{mahan2011condensed,zhou_perturbo_2021} for a Fermi energy 100~meV above the Dirac point. 
The resistivity curves for the two methods are in close agreement for all systems, including the overall magnitude and key features in the temperature dependence, such as the $T$-linear resistivity in MLG, activation of LB phonons in the TBG structures, and activation of optical phonons around 300 K. 
\\
\indent
\textit{Small Angle TBG.\textemdash}
The efficiency of our approach enables the study of low-angle TBG structures with several thousand atoms. We first calculate the temperature and filling dependent resistivity in 2.0\degree~TBG (3268-atom unit cell) and compare it with experiments~\cite{chung_transport_2018} in Fig.~\ref{fig:2deg}. 
Our calculations capture key trends observed in experiments on 2.0\degree~TBG, predicting resistivity values in very good agreement with experiments below 100 K and correctly describing the dependence of the resistivity on the electron density and band filling. Above 100 K, our calculations  underestimate the temperature dependence, leading to a resistivity at 300~K lower by approximately a factor of two relative to experiment.
We attribute this discrepancy to the known underestimation of e-ph coupling for optical modes in graphene within DFT~\cite{attaccalite_doped_2010}, which results in a similar discrepancy in first-principles resistivity calculations in MLG~\cite{sohier_phononlimited_2014,chen_intrinsic_2008, efetov_controlling_2010}. Note also that there is significant variation in the available experimental data. For example, Ref.~\cite{polshyn_large_2019} reports a 2\degree~TBG sample with a 300 K resistivity of over twice that of Ref.~\cite{chung_transport_2018} and near-$T$-linear temperature dependence.
\\
\indent  
We analyze the filling dependence of the resistivity in Fig.~\ref{fig:2deg}(b), which can be qualitatively compared to longitudinal resistance measurements in Ref.~\cite{chung_transport_2018}. Consistent with experiment, we find a greater resistivity below half filling than above, together with maxima associated with the positions of the van Hove singularities (VHSs) in the upper and lower moir\'e bands, and peaks near $n \!=\! \pm 10^{13}$ cm$^{-2}$ corresponding to completely filled or empty moir\'e bands, respectively. These peaks are smaller in our calculation than in experiment, a trend we attribute to the smaller band gaps between the moir\'e and dispersive bands in our calculations.
\\
\indent 
\textit{Twist angle dependence.\textemdash}
Next, we study the resistivity as a function of twist angle, reporting results for seven twist angles in Fig.~\ref{fig:twist-angle-all}(a). %
Our results reveal a substantial increase in resistivity with decreasing twist angles, with 1.61\degree~TBG exhibiting an approximately 40 times greater resistivity than 13.2\degree~TBG at 100 K. This trend is due primarily $-$ but not entirely $-$ to a decrease in band velocity with twist angle (see end matter). 
We also examine the twist angle dependence of the Eliashberg e-ph coupling constant $\lambda$~\cite{mahan2011condensed} [see Fig.~\ref{fig:twist-angle-all}(b)]. 
We observe a pronounced decrease in $\lambda$ with twist angle from 13.2\degree~to 2.65\degree, followed by a moderate increase at lower twist angles. The dependence of $\lambda$ on twist angle and Fermi energy closely follows the electronic density of states (DOS) [see Supplemental Material (SM)~\cite{supplemental_material}], indicating that the e-ph coupling strength depends only weakly on twist angle down to 1.61\degree.\\
\hspace{2pt}\textit{Analysis of transport trends.\textemdash}
We analyze in detail the dependence of the resistivity on temperature and electronic filling for three representative twist angles: 13.2\degree, a large angle where the layers are nearly decoupled, 3.15\degree, an intermediate value where the bands are substantially flatter but still have an energy scale several times larger than acoustic phonons, and 1.61\degree, a small twist angle where the energy scales of flat-band electrons and acoustic phonons become comparable. The band structures for these twist angles, shown in Fig.~\ref{fig:transport-twist-angle}(a)-(c) respectively, exhibit this evolution of the electronic energy scale.\\
\indent
For the same sequence of twist angles, the resistivity as a function of temperature and electronic filling are shown in Fig.~\ref{fig:transport-twist-angle}(d)-(f) and Fig.~\ref{fig:transport-twist-angle}(g)-(i), respectively. For the largest twist angle considered, 13.2\degree, the resistivity of TBG resembles that of MLG. 
The temperature dependence is approximately linear over the range of 50-200 K [Fig.~\ref{fig:transport-twist-angle}(d)], with some deviation due to LB phonons, as reported previously~\cite{gao_firstprinciples_2024}. The variation of the resistivity with electronic filling [Fig.~\ref{fig:transport-twist-angle}(g)] is small over the range of -0.5 to 0.5 electrons per unit cell, where the main trend is a decrease in resistivity with increasing filling due changes in the band velocity. 
As in MLG, this transport behavior is related to the Dirac band dispersion: the band velocity varies slowly over the energy range associated with transport, and the electronic density of states increases linearly with Fermi energy (see further analysis in the SM~\cite{supplemental_material}). 
\\
\indent
Some of these features are also present in the resistivity of 3.15\degree~TBG, but significant deviations emerge [Figs.~\ref{fig:transport-twist-angle}(e) and \ref{fig:transport-twist-angle}(h)]. The resistivity remains approximately $T$-linear over the 50-200 K temperature range for fillings within $\pm$1-2 electrons per unit cell, but it exhibits different trends when the moir\'e bands are close to empty or full. 
This behavior arises from features in the electronic structure of the moir\'e bands, including VHSs around $\pm150$ meV that produce maxima in the density of states and minima in the band velocity. These sharp features modify the temperature dependence of the resistivity when their energy scale becomes comparable to $k_BT$ and the phonon energy. 
\\
\indent
At the smallest twist angle considered, 1.61\degree, the resistivity deviates significantly from the Dirac-dispersion behavior, and the temperature dependence of the resistivity becomes strongly non-linear [Fig.~\ref{fig:transport-twist-angle}(f)]. For electronic fillings near $\pm$1 electron per unit cell, a nearly linear regime appears below $\sim$100~K, followed by a pronounced deviation at higher temperatures. 
In 1.61\degree~TBG, features in the electronic structure occur on an energy scale of $\sim$20 meV, so the electronic states contributing to transport vary with temperature, leading to a more complex temperature dependence. 
When the moir\'e bands are completely filled or empty, the system becomes insulating due to the small gaps between the moir\'e and dispersive bands, as observed experimentally at similar twist angles~\cite{chung_transport_2018}. The filling dependent resistivity [Fig.~\ref{fig:transport-twist-angle}(i)] reflects the low moir\'e energy scale, with peaks appearing near the VHSs at 50 K. 
These features broaden as $k_BT$ becomes comparable to the electronic energy scale, and the dominant feature in the filling dependence becomes a maximum around -1 electron per unit cell, corresponding to a Fermi energy near the lower VHS. 
\\
\indent 
Taken together, our results indicate that the phonon-limited Boltzmann transport picture remains valid for twist angles down to $\sim$1.5\degree~over a relatively wide temperature range. 
The comparison with the experiments of Ref.~\cite{chung_transport_2018} in Fig.~\ref{fig:2deg}(a) supports this conclusion. 
In addition, measurements of 1.59\degree~TBG at half filling of the lower moir\'e bands (-2 electrons per unit cell) report a $T$-linear resistivity with a value of 1.6 k$\Omega$ at 300 K and a slope of $\sim$5 $\Omega / \mathrm{K}$. Our calculations for 1.61\degree~TBG predict a $T$-linear regime below 100 K and a slope of 2.5 $\Omega / \mathrm{K}$ at 300~K, in agreement with the experiment within a factor of two. Our computed resistivity exhibits a weaker temperature dependence than experiment above 100~K for both 1.61\degree~and 2.0\degree. As discussed above, this discrepancy originates from the known underestimation of optical-phonon coupling in DFT, previously identified in studies of MLG, which limits the accuracy of the predicted resistivity at higher temperatures. 
\\
\indent
\textit{Discussion.\textemdash}
Because of its rich physics, modeling interactions in low-angle TBG remains challenging. This Letter describes e-ph interactions within the adiabatic picture and transport using the Boltzmann equation. Electronic correlation and screening effects are described using semilocal DFT and DFPT, which provide the parametrization of our e-ph potential. Going beyond this standard framework requires addressing several points. 
\\
\indent
First, correlation effects become dominant and strongly twist-angle dependent near the magic angle, where TBG behaves as a strongly correlated material~\cite{cao_correlated_2018, xie_nature_2020, choi_interaction_2021}, with substantial changes expected for e-ph interactions~\cite{huang_electronphonon_2003, abramovitch_respective_2024, mandal_strong_2014, li_electronphonon_2019}. A description of nonlocal e-ph interactions using the GW method and of local electronic interactions using dynamical mean field theory would be highly desirable in this regime, where building tight-binding-like models is similarly challenging~\cite{koshino_maximally_2018, goodwin_twist_2019, koshino_maximally_2018,carr_derivation_2019}. Second, an improved description of transport would need to build on these advances. 
For example, electronic correlations induce filling-dependent band flattening in low-angle TBG~\cite{choi_interaction_2021}, enhancing the resistivity through reduced band velocity and modifying its temperature and filling dependence. In addition, electron-electron scattering is expected to play an important role in the resistivity of magic-angle TBG~\cite{shilov_interaction_2025}. Describing this physics requires an improved framework for both electronic interactions and transport. 
Finally, the adiabatic approximation breaks down near the magic angle, where electronic and phonon energy scales become comparable. This hinders direct application of the Boltzmann equation for transport and the Migdal–Eliashberg formalism for superconductivity at the magic angle. These open questions offer exciting directions for future work. 
\vspace{2pt}\\
\indent 
\textit{Conclusion.\textemdash}
In summary, we have developed calculations of e-ph coupling and electronic transport at the moir\'e scale in TBG using an atomistic e-ph interaction potential containing Holstein and Peirls terms. 
Our study of transport across a range of twist angles reveals that the resistivity increases in magnitude and gains additional features in the temperature and filling dependence as the moir\'e energy scale decreases, highlighting the rich interplay of electron and phonon energy scales in low-angle TBG. Comparisons with experiments indicate that e-ph interactions limit the resistivity in TBG for angles as small as $\sim$1.61\degree.  
Our quantitative framework for e-ph coupling provides a starting point for investigating the effects of correlations on e-ph interactions in TBG, with the ultimate goal of understanding e-ph interactions near the magic angle and their role in TBG superconductivity and phase diagram~\cite{wu_theory_2018, choi_strong_2018}.
Beyond TBG, our scalable e-ph calculations provide access to e-ph coupling and associated properties in large systems, opening avenues for modeling disparate materials such as heterojunctions, superlattices, quasicrystals, and device-scale structures. This framework can also be integrated with machine learning workflows, opening opportunities for high-throughput studies of e-ph interactions.\vspace{6pt}\\
\vfill
\indent 
\textit{Acknowledgements.\textemdash} 
D.J.A. acknowledges Kiran Shila for helpful discussions. 
D.J.A. is partially supported by the National Science Foundation Graduate Research Fellowship under Grant No. 2139433.  
This research used resources of the National Energy Research Scientific Computing Center, a DOE Office of Science User Facility supported by the Office of Science of the U.S. Department of Energy under Contract No. DE-AC02-05CH11231 using NERSC award NERSC DDR-ERCAP0026831.
\clearpage
\bibliography{bibliography}
\newpage

\setcounter{section}{0}
\setcounter{figure}{0}
\end{document}